\documentclass[aps, showpacs,nofootinbib,floatfix]{revtex4}

\usepackage{amssymb}
\usepackage{amsmath}
\usepackage{graphicx}

\begin{document}

\title{Cosmic Constraint to DGP Brane Model: Geometrical and Dynamical Perspectives}

\author{Lixin Xu\footnote{Corresponding author}}
\email{lxxu@dlut.edu.cn}
\author{Yuting Wang}

\affiliation{Institute of Theoretical Physics, School of Physics \&
Optoelectronic Technology, Dalian University of Technology, Dalian,
116024, P. R. China}

\begin{abstract}
In this paper, the Dvali-Gabadadze-Porrati (DGP) brane model is
confronted by current cosmic observational data sets from
geometrical and dynamical perspectives. On the geometrical side, the
recent released Union2 $557$ of type Ia supernovae (SN Ia), the
baryon acoustic oscillation (BAO) from Sloan Digital Sky Survey and
the Two Degree Galaxy Redshift Survey (transverse and radial to
line-of-sight data points), the cosmic microwave background (CMB)
measurement given by the seven-year Wilkinson Microwave Anisotropy
Probe observations (shift parameters $R$, $l_a(z_\ast)$ and redshift
at the last scatter surface $z_\ast$), ages of high redshifts
galaxies, i.e. the lookback time (LT) and the high redshift Gamma
Ray Bursts (GRBs) are used. On the dynamical side, data points about
the growth function (GF) of matter linear perturbations are used.
Using the same data sets combination, we also constrain the flat
$\Lambda$CDM model as a comparison. The results show that current
geometrical and dynamical observational data sets much favor flat
$\Lambda$CDM model and the departure from it is above
$4\sigma$($6\sigma$) for spatially flat DGP model with(without) SN
systematic errors. The consistence of growth function data points is
checked in terms of relative departure of redshift-distance
relation.
\end{abstract}



\maketitle

\section{Introduction}

Understanding the current accelerated expansion of our universe has
become one of the most important issues of modern cosmology
\cite{ref:Riess98,ref:Perlmuter99}. However, as so far, we still
know little about the nature of current accelerated expansion. In
general, from the phenomenological points of view, the possible
models can be classified by the form of Friedmann equation. One is
\begin{equation}
H^2=f(\rho),
\end{equation}
where the extra energy component(s) is(are) added on the right hand
of Einstein's equations, and $f(\rho)$ is a function of energy
density $\rho$ which can be composed from dark matter and extra
energy components. Here $H$ is the Hubble parameter. The other is
\begin{equation}
g(H^2)=\rho,
\end{equation}
where the energy component(s) is (are) invariant and the {\it
Gravity Theory} or equivalently the relation of Hubble parameter $H$
with the conventional matter (dark matter at late time) can be
altered, $g(H^2)$ is a function of $H^2$ or $H$. For recent reviews
about dark energy and an accelerated expansion universe, please see
\cite{ref:reviews}.

In this paper, we will consider a leading modified gravity model,
Dvali-Gabadadze-Porrati (DGP) brane model \cite{ref:DGP}, for the
review, please see \cite{ref:DGPreview}. In DGP brane model, a
tensionless four-dimensional brane (a hypersurface with a vanishing
cosmological constant) is embedded in a five dimensional bulk which
is a flat Minkowski space-time, where the gauge forces are confined
on the brane and gravity can propagate in all dimensions freely.
Below the crossover scale $r_c$, the gravity appears
four-dimensional. However, above the scale $r_c$ the gravity can
leak into the extra dimension and make the conventional
four-dimensional gravity altered. It is due to the leakage of
gravity, the current universe appears an accelerated expansion
phase.

In the past years, the DGP model has been constrained by cosmic
observational data sets, for recent constrained results, please see
\cite{ref:DGPGc} in geometrical side and \cite{ref:LDGPDc} in
dynamical side. Importantly, Lombriser {\it et.al.}
\cite{ref:DGPFang} constrained DGP brane model exhaustively by
adopting a parameterized post-Friedmann description of gravity,
where all of the CMB data, including the largest scales, and its
correlation with galaxies in addition to the geometrical constraints
from supernovae distances and the Hubble constant were utilized. Of
course, it is more important and necessary to constrain a
cosmological model by using all of CMB data and its correlation with
large scale structure, say galaxy. However, it is much complicated
when we have to resort to some kinds of Boltzmann-codes, say the
famous {\bf CMBfast} and {\bf CAMB}. It would be interesting, if
this complication is avoided, and a relative stronger prediction is
arrived. In this paper, we will investigate this kind of possibility
by combining the geometrical and dynamical perspectives without the
above mentioned complication. Via this combination, the constraint
becomes much tight and efficient because it relies on both gravity
theory and background geometry. On the geometrical side, we will
consider the luminosity-distance relation of SN and GRB, the
standard rulers from BAO, x-ray gas fraction and CMB, also the
lookback time of high redshift galaxy. For the dynamical
perspective, we mainly consider the growth function $\delta(z)\equiv
\frac{\delta\rho}{\rho}(z)$ of the linear matter density contrast as
a function of redshift. As well known, the data points of growth
function were obtained with the help of $\Lambda$CDM model and the
reliability will be reduced when it is used to constrain other
cosmological models, however this problem can be evaded in the DGP
model if it has similar expansion history to that of $\Lambda$CDM
\cite{ref:LDGPDc,ref:DGPwei}. In fact, the use of redshift-distance
relation of $\Lambda$CDM (with $\Omega_m=0.25$ for the data point at
$z = 0.77$ and $\Omega_m=0.30$ for the remained five data points)
makes the the data points of growth function weak to constrain other
cosmological model \cite{ref:gfweak}. In \cite{ref:DGPwei}, to use
these data points to find the growth index of DGP model, the DGP
model parameters were selected properly to make redshift-distance
relation be the same as that of $\Lambda$CDM model. In fact, we can
firstly neglect this weak point and just use these data points as
dynamical constraint beyond $\Lambda$CDM model. After it was done,
we can check the consistence via investigating the possible
departure from $\Lambda$CDM model with the best fit values of model
parameters. If the departure is large, the data points must be
discarded and the data fitting is not reliable. In fact, one will
find that the redshift-distance relations of $\Lambda$CDM and DGP
model are almost the same and the relative departure from
$\Lambda$CDM model is up to $10\%$ in terms of $H_0 r(z)$ when $\Omega_m$ varies in
the range $[0.28,0.32]$, where $\Omega_m=0.30$ is fixed in $\Lambda$CDM model. When
the values of $\Omega_m$ increase, the relative departure becomes small.
And the departure is
smaller than the
errors of growth function data points. Based on this point, the data
points of growth function can be used safely to constrain DGP model.
Also, when the data points of growth function are used to confront
other cosmological models, the consistence must be checked.

For the DGP model, the modified Friedmann equation is given as
\cite{ref:DGPFE}
\begin{equation}
H^2+\frac{k}{a^2}=\left[\sqrt{\frac{\rho}{3M^2_{pl}}+\frac{1}{4r_c^2}}+\frac{1}{2r_c}\right]^2
\end{equation}
where $r_c=M^2_{pl}/2M^3_5$ is the so-called crossover scale,
$M_{pl}$ and $M_5$ are four- and five-dimensional reduced Planck
mass respectively, $\rho$ is the cosmological fluid which includes
conventional matter contents and radiation, $k=0,\pm 1$ is three
dimensional spatial curvature factor. In terms of the dimensionless
density parameters, the Friedmann equation can be rewritten as
\begin{equation}
E^2(z)=\Omega_{k}(1+z)^2+\left(\sqrt{\Omega_{m}(1+z)^3+\Omega_{r}(1+z)^4+\Omega_{r_c}}+\sqrt{\Omega_{r_c}}\right)^2,
\end{equation}
where $E^2(z)=H^2(z)/H^2_0$ and $\Omega_{r_c}=1/(4r^2_c H^2_0)$ is a
constant which respects to the constraint equation
\begin{equation}
\Omega_{k}+\left(\sqrt{\Omega_{m}+\Omega_{r}+\Omega_{r_c}}+\sqrt{\Omega_{r_c}}\right)^2=1.
\end{equation}
For the spatially flat case ($\Omega_{k}=0$), the above equation
reduces to $\Omega_{r_c}=(1-\Omega_m-\Omega_r)^2/4$.

\section{Method and Results}

In our calculations, we have taken the total likelihood function
$L\propto e^{-\chi^2/2}$ to be the products of the separate
likelihoods of SN (with and without systematic errors), BAO, CMB,
GRBs, CBF, LT and GF. Then we get $\chi^2$
\begin{eqnarray}
\chi^2=\chi^2_{SN}+\chi^2_{BAO}+\chi^2_{CMB}+\chi^2_{GRBs}+\chi^2_{CBF}+\chi^2_{LT}+\chi^2_{GF},
\end{eqnarray}
where the separate likelihoods of SN, BAO, CMB, GRBs, CBF, LT, GF
and the current observational datasets used in this paper are shown
in the Appendix \ref{app:obervations}.

In our analysis, we perform a global fitting to determine the
cosmological parameters using the Markov Chain Monte Carlo (MCMC)
method. The MCMC method is based on the publicly available {\bf
CosmoMC} package \cite{ref:MCMC} and the {\bf modified CosmoMC}
package \cite{ref:0409574,ref:07060033,ref:modifiedMCMC}, including
the X-ray cluster gas mass fraction. For our models we have modified
these packages to add some subroutines, for example the code
calculating the growth function likelihood etc. The following basic
cosmological parameters ($\Omega_bh^2$, $\Omega_ch^2$) are varying
with top-hat priors: the physical baryon density
$\Omega_{b}h^2\in[0.005,0.1]$, the physical dark matter energy
density $\Omega_{c}h^2\in[0.01,0.99]$. And, in the data fitting
process another seven parameters ($K, \eta, \gamma, b_0, \alpha_b,
s_0, \alpha_s$) included in the X-ray gas mass fraction $f_{gas}$
are treated as free parameters. As a byproduct the best fitting
values of these parameters are obtained. And, these values can also
be taken accounted as a check of data fitting.

\begin{figure}[!htbp]
\includegraphics[width=8cm]{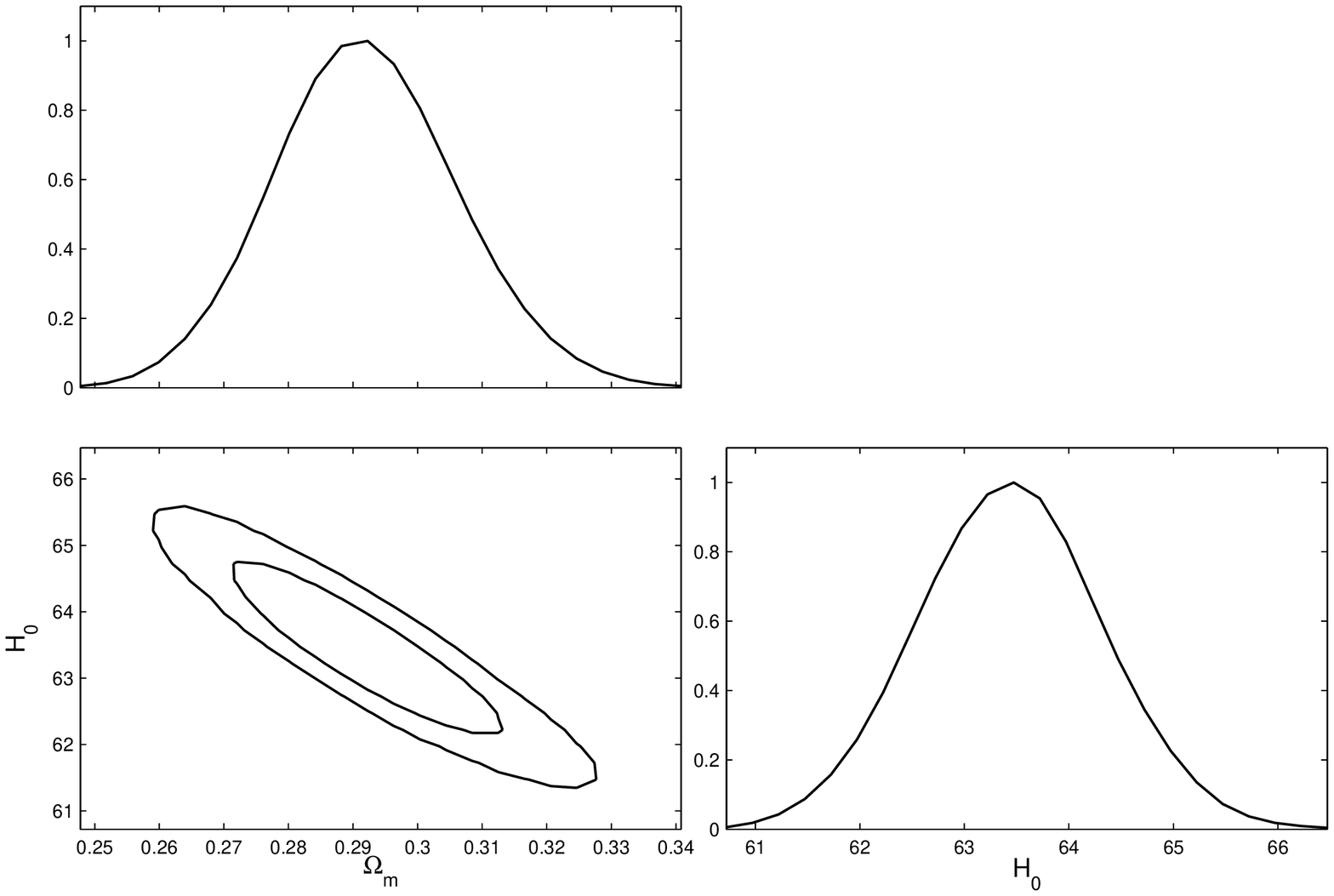}
\includegraphics[width=8cm]{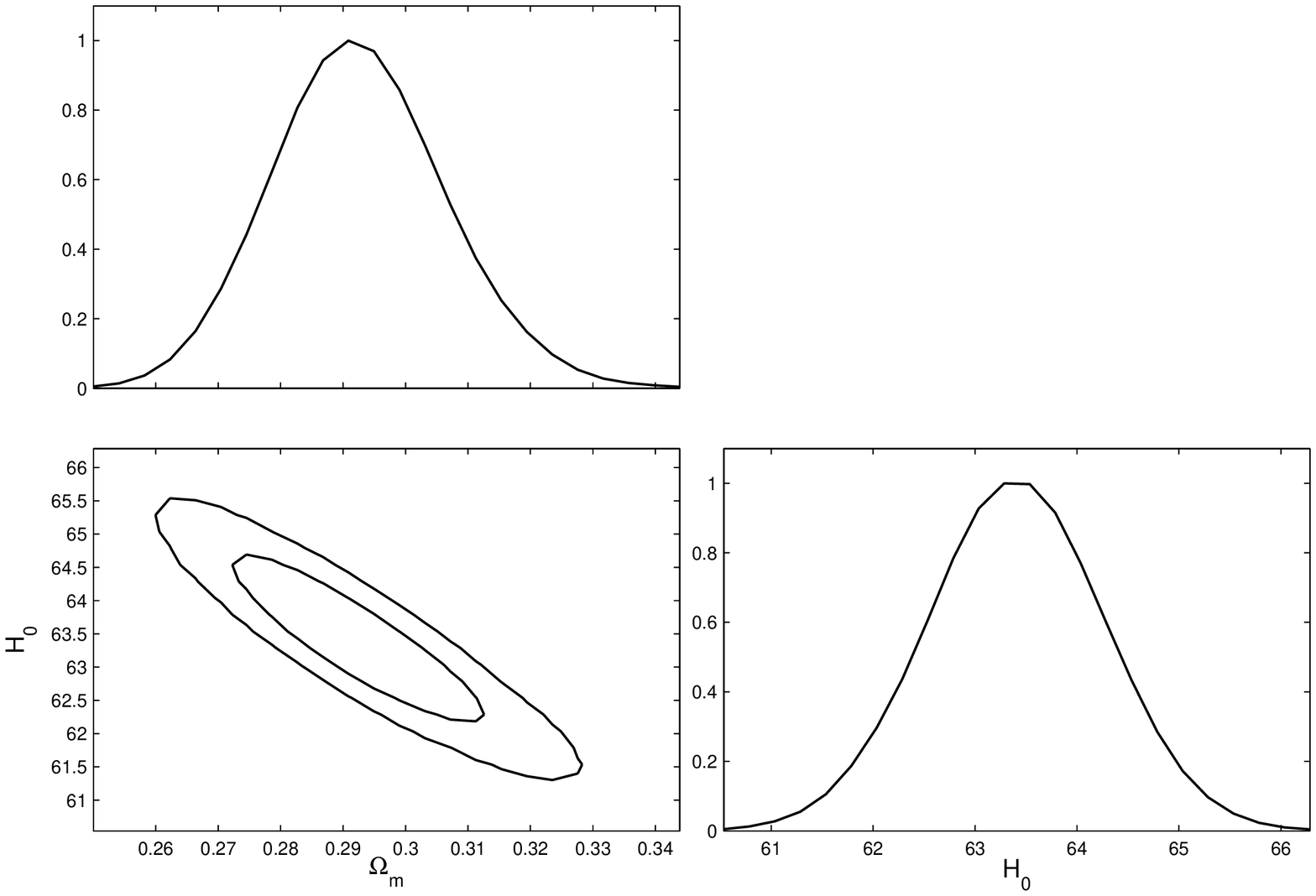}
\caption{The 2-D contours with $1\sigma$, $2\sigma$ regions and 1-D
marginalized distribution of $\Omega_{m}$, $H_0$. Left Panel:
$\chi^2_{min}=663.561$, $\Omega_{m}=0.285^{+0.0470}_{-0.0286}$ and
$H_0=63.754^{+2.142}_{-2.575}$ with $5$ growth function data points,
$z=3.0$ is removed. Right Panel: $\chi^2_{min}=673.641$,
$\Omega_{m}=0.297^{+0.0367}_{-0.0391}$ and
$H_0=63.121^{+2.629}_{-2.0471}$ with all $6$ growth function data
points included.}\label{fig:DGP}
\end{figure}

\begin{figure}[!htbp]
\includegraphics[width=12cm]{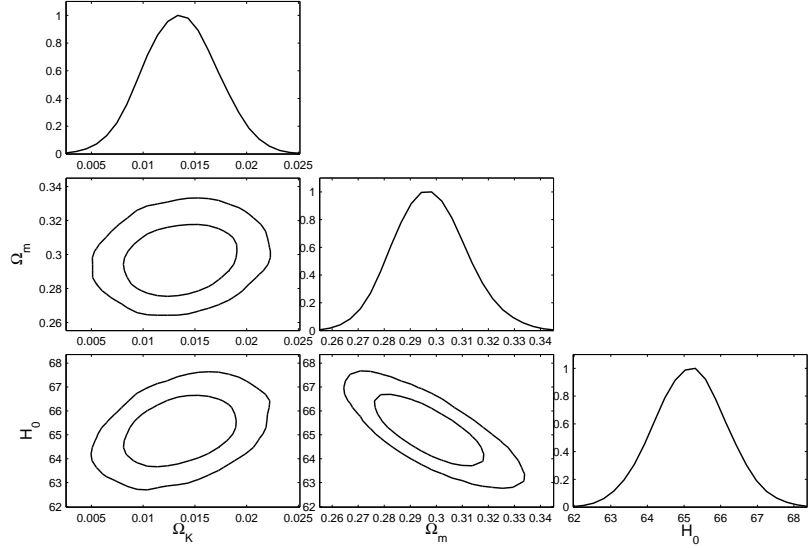}
\caption{The 2-D contours with $1\sigma$, $2\sigma$ regions and 1-D
marginalized distribution of $\Omega_{m}$, $H_0$ and $\Omega_{k}$ in
non-flat DGP model without $z=3.0$ data
point.}\label{fig:DGPnonflat}
\end{figure}

From the observational results listed in Tab. \ref{tab:fobs}, one
has noticed that the datum at $z=3.0$ is odd in some degree for its
value is above $1$. So in data fitting process, we have considered
two cases with and without inclusion of $z=3.0$ data point. The
constrained results are shown in Tab. \ref{tab:results} and Fig.
\ref{fig:DGP} (with SN systematic errors). From Tab.
\ref{tab:results}, one can easily see that the value of $\chi^2_{min}$
is smaller and the constraint is tighter without $z=3.0$ data point
than that with it. Under this status, one would constrain a
cosmological model without the inclusion of $z=3.0$ data point. For
the non-flat DGP model, we summarize the model parameters values in
Tab. \ref{tab:resultsnonflat} and Fig. \ref{fig:DGPnonflat} (with SN
systematic errors). As a comparison, using the same combination
(with SN systematic errors) to the flat $\Lambda$CDM model, we have
the results $\chi^2_{min}=626.057$,
$\Omega_{m}=0.281^{+0.0300}_{-0.0360}$ and
$H_0=69.658^{+3.0176}_{-2.142}$. One can see that the best fit
values of $\Omega_m$ and $1\sigma$ errors of model parameters are
enlarged when SN systematic errors are included. As contrasts, the
current Hubble values and $\chi^2_{min}$ are lower in the case of SN
systematic errors included.

\begin{table}
\begin{center}
\begin{tabular}{cc|    cc    cc|  cc |  cc }
\hline\hline Datasets & & Parameters & & $\chi^2_{min}/{\rm d.o.f}$&
& $\Omega_{m}$ & & $H_0$
\\ \hline
GF5 && 9 &        & 697.342/645&
                   & $0.265^{+0.0291}_{-0.0302}$ &  & $65.219^{+2.190}_{-1.956}$ \\
GF6  && 9 &        & 707.750/646&
                   & $0.266^{+0.0298}_{-0.0304}$ &  & $65.0873^{+2.413}_{-1.985}$ \\
GF5(SN Sys.) && 9 &        & 663.561/645&
                   & $0.285^{+0.0470}_{-0.0286}$ &  & $63.754^{+2.142}_{-2.575}$ \\
GF6(SN Sys.)  && 9 &        & 673.641/646&
                   & $0.297^{+0.0367}_{-0.0391}$ &  & $63.121^{+2.629}_{-2.0471}$ \\
\hline\hline
\end{tabular}
\caption{The results of $\chi^2_{min}$, $\Omega_{m}$ and $H_0$ with
$1\sigma$ regions are listed. GF5 denotes $5$ growth function data
points are used without $z=3.0$ data point. And, GF6 denotes the
corresponding $6$ data points case. SN Sys. denotes the results with
SN systematic errors. ${\rm d.o.f}$ denotes the degrees of
freedom.}\label{tab:results}
\end{center}
\end{table}

\begin{table}
\begin{center}
\begin{tabular}{cc|    cc    cc|  cc |  cc | cc}
\hline\hline Datasets & & Parameters & & $\chi^2_{min}/{\rm d.o.f}$&
& $\Omega_{m}$ & & $H_0$ & & $\Omega_{k}$
\\ \hline
non-flat && 10 &        & 684.251/644&
                   & $0.270^{+0.0278}_{-0.0323}$ &  & $66.798^{+2.85}_{-2.483}$ & & $0.0123^{+0.00789}_{-0.00993}$\\
non-flat(SN Sys.) && 10 &        & 647.543/644&
                   & $0.295^{+0.0419}_{-0.0345}$ &  & $65.519^{+2.490}_{-2.974}$ & & $0.0153^{+0.00777}_{-0.01052}$\\
\hline\hline
\end{tabular}
\caption{The results of $\chi^2_{min}$, $\Omega_{m}$, $H_0$ and
$\Omega_{k}$ without $z=3.0$ data point in non-flat DGP model. SN
Sys. denotes the results with SN systematic errors. ${\rm d.o.f}$
denotes the degrees of freedom.}\label{tab:resultsnonflat}
\end{center}
\end{table}

\begin{figure}[!htbp]
\includegraphics[width=12cm]{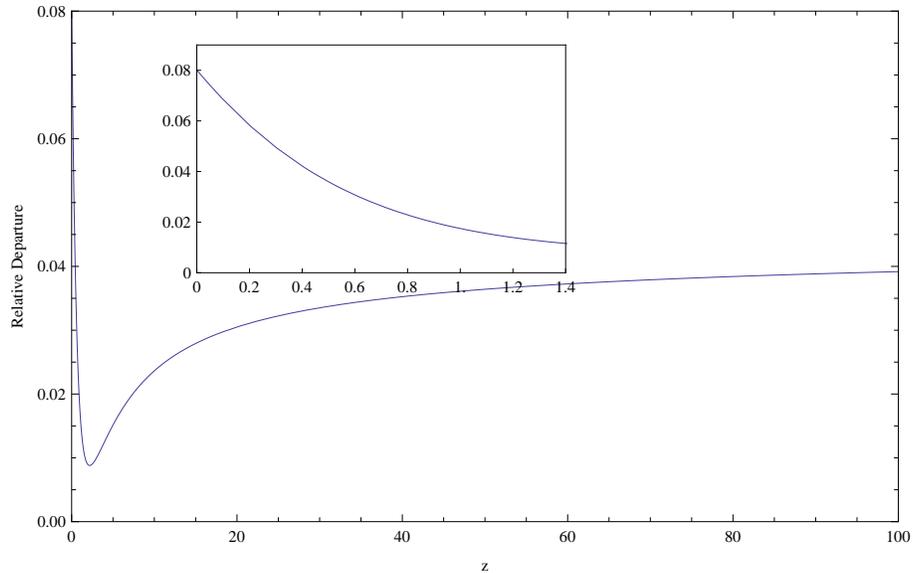}
\caption{The relative departure of redshift-distance $r(z)$ of DGP
model with the best fit values of model parameters from $\Lambda$CDM
model with $\Omega_m=0.30$ and $H_0=69.658$.}\label{fig:departure}
\end{figure}

\begin{figure}[!htbp]
\includegraphics[width=12cm]{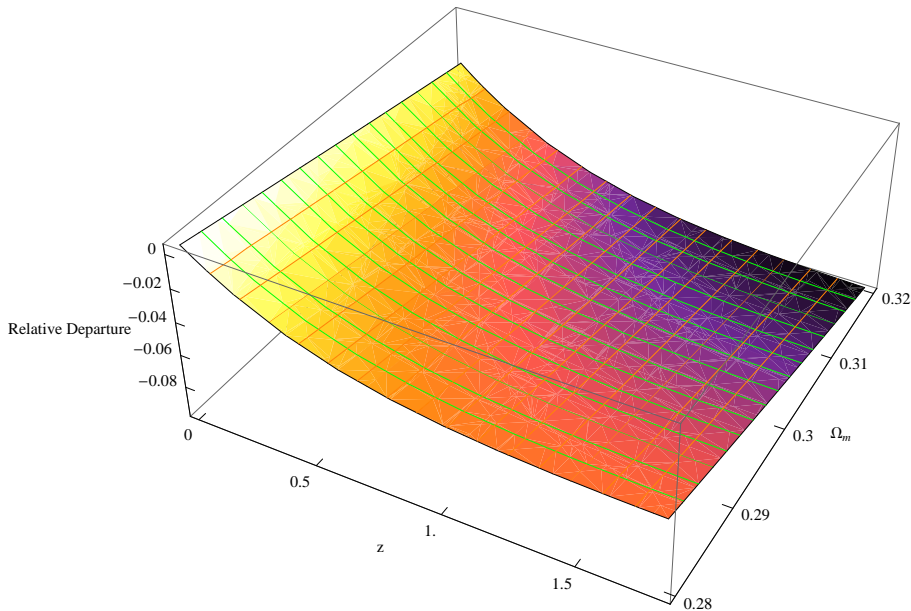}
\caption{The relative departure of redshift-distance in terms of $H_0r(z)$ of DGP model
from $\Lambda$CDM model with redshift and $\Omega_m$ where $\Omega_m=0.3$ is
fixed in $\Lambda$CDM model.}\label{fig:departure3d}
\end{figure}

Now, it's time to check the consistence of our treatment of the
points of growth function. We calculate the relative departure of
comoving distance
\begin{equation}
{\rm rel.dep.}= \frac{r_{DGP}(z)-r_{\Lambda CDM}(z)}{r_{\Lambda
CDM}(z)}
\end{equation}
where $\Omega_m=0.30$ is fixed in $\Lambda$CDM model. The result is
shown in Fig. \ref{fig:departure}. In fact, one can check the
consistence via the relative departure in terms $H_0 r(z)/c$ with the variables of $z$ and
$\Omega_m$, the corresponding result is shown
in a 3D plot, see Fig. \ref{fig:departure3d}. Clearly, the relative
departures are up to $10\%$ for neglecting the uncertainty of $H_0$.
And, with the increasing value of $\Omega_m$, the departure is amplifying. In fact, as shown in Fig.
\ref{fig:departure}, the departure is just up to $8\%$ when the best fit values of model parameters are
adopted.

\section{Conclusion and Discussion}

In summary, in this paper we have performed a global fitting on the
cosmological parameters in both the flat DGP model and the non-flat
DGP model by using a completely consistent analysis from the
geometrical and dynamical perspectives. On the geometrical side, the
X-ray gas mass fraction observation, type Ia supernovae data from
Union2 set, transverse and radial baryon acoustic oscillations data
from SDSS, the measurement data on current Cosmic Microwave
Background from the seven-year WMAP observations, the lookback time
data derived from the ages of galaxy and clusters and the high
redshift gamma ray bursts. On the dynamical side, the data points
about the growth function of matter linear perturbations are
included. The constrained results are shown in Tab.
\ref{tab:results} for the flat case and Tab.
\ref{tab:resultsnonflat} for the non-flat case. The results show
that the $z=3.0$ data point of growth function is odd in some
degrees. So, in the future work when the observational growth
function data points are used as cosmic constraint, this point would
be removed. As a  comparison, using the same data points
combination, the flat $\Lambda$CDM model was constrained, where we
have $\chi^2_{min}=626.057$, $\Omega_{m}=0.281^{+0.0300}_{-0.0360}$
and $H_0=69.658^{+3.0176}_{-2.142}$. Clearly, one can easily find
that $\Delta\chi^2=37.504$ which in terms of $\sigma$ distance is
$4.25$ for $9$ parameters via the formula
$1-\Gamma(\nu/2,\Delta\chi^2/2)/\Gamma(\nu/2)={\rm
Erf}(d_{\sigma}/\sqrt{2})$, where $\nu=9$ is the number of free
model parameters. It means that current geometrical and dynamical
observational data sets favor flat $\Lambda$CDM model more above
$4\sigma$ than that for spatially flat DGP model. This conclusion is
consistent and comparable with that of \cite{ref:DGPFang}. When the
same process of data fitting is implemented without SN systematic
errors, the departure from $\Lambda$CDM model can be improved to
about $6\sigma$ for the flat DGP model. To check the consistence of
growth function data points, the relative departure of DGP model
with best fit model parameters from $\Lambda$CDM model with fixed
$\Omega=0.30$ is up to $8\%$. In general, the relative departure in
terms of $H_0r(z)$ is up to $10\%$ when $\Omega_m$ varies in the
range $[0.28,0.32]$ and $\Omega_m=0.30$ is fixed in $\Lambda$CDM
model. With this observation, we can say that growth function data
points can be used to constrain DGP model. So far, we have seen that
it is possible to give some strong prediction without having to
resort to the complicated modification of the Boltzmann-codes. At
least, for flat DGP model, it is possible. It is because that the
discrepancy will be enlarged when more observational data points are
included. We expect this work can shed light on distinguish dark
energy models. At last, we are pleasant to give some comments on the
growth function data points used to constrain other cosmological
models. The key point is the dependence of distance-redshift
relation of spatially flat $\Lambda$CDM model with fixed value of
$\Omega_m=0.30$. Of course, one can firstly neglect this fact and do
data fitting as what we have done in this paper. Then, one has to
check the consistence. In fact, we can get around this problem in a
different way. It is the full consideration of distance-redshift
relation and introduction of some kinds of uncertainties in terms of
redshift and growth function data points, put in another words, via
the possible introduction of extra errors on redshifts and growth
function data points. Of course, it is out the range of this paper.
In the future work, this possibility will be researched.

\acknowledgements{This work is supported by NSF (10703001), SRFDP
(20070141034) of P.R. China. We thank Prof. Deepak Jain and Dr. Yun
Chen for the correspondence on lookback time data set. We appreciate
the anonymous referee's invaluable help to improve this work.}

\appendix

\section{Cosmological Constraints Methods}\label{app:obervations}

\subsection{Type Ia Supernovae constraints}

Recently, SCP (Supernova Cosmology Project) collaboration released
their Union2 dataset which consists of 557 SN Ia \cite{ref:SN557}.
The distance modulus $\mu(z)$ is defined as
\begin{equation}
\mu_{th}(z)=5\log_{10}[\bar{d}_{L}(z)]+\mu_{0},
\end{equation}
where $\bar{d}_L(z)$ is the Hubble-free luminosity distance $H_0
d_L(z)/c=H_0 d_A(z)(1+z)^2/c$, with $H_0$ the Hubble constant,
defined through the re-normalized quantity $h$ as $H_0=100 h~{\rm km
~s}^{-1} {\rm Mpc}^{-1}$, and $\mu_0\equiv42.38-5\log_{10}h$. Where
$d_L(z)$ is defined as
\begin{equation}
d_L(z)=(1+z)r(z),\quad r(z)=\frac{c}{H_0\sqrt{|\Omega_{k}|}}{\rm
sinn}\left[\sqrt{|\Omega_{k}|}\int^z_0\frac{dz'}{E(z')}\right]
\end{equation}
where $E^2(z)=H^2(z)/H^2_0$. Additionally, the observed distance
moduli $\mu_{obs}(z_i)$ of SN Ia at $z_i$ is
\begin{equation}
\mu_{obs}(z_i) = m_{obs}(z_i)-M,
\end{equation}
where $M$ is their absolute magnitudes.

For the SN Ia dataset, the best fit values of the parameters $p_s$
can be determined by a likelihood analysis, based on the calculation
of
\begin{eqnarray}
\chi^2(p_s,M^{\prime})&\equiv& \sum_{SN}\frac{\left\{
\mu_{obs}(z_i)-\mu_{th}(p_s,z_i)\right\}^2} {\sigma_i^2}  \nonumber\\
&=&\sum_{SN}\frac{\left\{ 5 \log_{10}[\bar{d}_L(p_s,z_i)] -
m_{obs}(z_i) + M^{\prime} \right\}^2} {\sigma_i^2}, \label{eq:chi2}
\end{eqnarray}
where $M^{\prime}\equiv\mu_0+M$ is a nuisance parameter which
includes the absolute magnitude and the parameter $h$. The nuisance
  parameter $M^{\prime}$ can be marginalized over
analytically \cite{ref:SNchi2} as
\begin{equation}
\bar{\chi}^2(p_s) = -2 \ln \int_{-\infty}^{+\infty}\exp \left[
-\frac{1}{2} \chi^2(p_s,M^{\prime}) \right] dM^{\prime},\nonumber
\label{eq:chi2marg}
\end{equation}
resulting to
\begin{equation}
\bar{\chi}^2 =  A - \frac{B^2}{C} + \ln \left( \frac{C}{2\pi}\right)
, \label{eq:chi2mar}
\end{equation}
with
\begin{eqnarray}
&&A=\sum_i^{SN} \frac {\left\{5\log_{10}
[\bar{d}_L(p_s,z_i)]-m_{obs}(z_i)\right\}^2}{\sigma_i^2},\nonumber\\
&& B=\sum_i^{SN} \frac {5
\log_{10}[\bar{d}_L(p_s,z_i)]-m_{obs}(z_i)}{\sigma_i^2},\nonumber
\\
&& C=\sum_i^{SN} \frac {1}{\sigma_i^2}\label{eq:SNABC}.
\end{eqnarray}
Relation (\ref{eq:chi2}) has a minimum at the nuisance parameter
value $M^{\prime}=B/C$, which contains information of the values of
$h$ and $M$. Therefore, one can extract the values of $h$ and $M$
provided the knowledge of one of them. Finally, note that the
expression
\begin{equation}
\chi^2_{SN}(p_s,B/C)=A-(B^2/C),\label{eq:chi2SN}
\end{equation}
which coincides to Eq. (\ref{eq:chi2mar}) up to a constant, is often
used in the likelihood analysis \cite{ref:smallomega,ref:SNchi2},
and thus in this case the results will not be affected by a flat
$M^{\prime}$ distribution. It worths noting that the results will be
altered without the systematic errors. In this work, two cases with
and without systematic errors are considered together. When the
systematic errors are included, the corresponding $A,B,C$ are
expressed as
\begin{eqnarray}
A&=&\sum_{i,j}^{SN}\left\{5\log_{10}
[\bar{d}_L(p_s,z_i)]-m_{obs}(z_i)\right\}\cdot C^{-1}_{ij}\cdot
\left\{5\log_{10}
[\bar{d}_L(p_s,z_j)]-m_{obs}(z_j)\right\},\nonumber\\
B&=&\sum_i^{SN} C^{-1}_{ij}\cdot \left\{5\log_{10}
[\bar{d}_L(p_s,z_j)]-m_{obs}(z_j)\right\},\nonumber \\
C&=&\sum_i^{SN} C^{-1}_{ii},\label{eq:SNsyserror}
\end{eqnarray}
where $C^{-1}$ is the inverse of covariance matrix with systematic
errors. For the details and covariance matrice, one can find them in
Ref. \cite{ref:SN557} and the web site
\footnote{http://supernova.lbl.gov/Union/}, where one can also find
the covariance matrix without systematic errors. Our form
(\ref{eq:SNABC}) is equivalent to (\ref{eq:SNsyserror}) when
$C^{-1}$ is the inverse of covariance matrix without systematic
errors.

\subsection{Baryon Acoustic Oscillation constraints}

The Baryon Acoustic Oscillations are detected in the clustering of
the combined 2dFGRS and SDSS main galaxy samples, and measure the
distance-redshift relation at $z = 0.2$. Additionally, Baryon
Acoustic Oscillations in the clustering of the SDSS luminous red
galaxies measure the distance-redshift relation at $z = 0.35$. The
observed scale of the BAO calculated from these samples, as well as
from the combined sample, are jointly analyzed using estimates of
the correlated errors to constrain the form of the distance measure
$D_V(z)$
\cite{ref:Okumura2007,ref:Percival2,ref:Eisenstein05,ref:Percival3}
\begin{equation}
D_V(z)=c\left(\frac{z}{\Omega_k
H(z)}\mathrm{sinn}^2[\sqrt{|\Omega_k|}\int_0^z\frac{dz'}{H(z')}]\right)^{1/3}.
\label{eq:DV}
\end{equation}
The peak positions of the BAO depend on the ratio of $D_V(z)$ to the
sound horizon size at the drag epoch (where baryons were released
from photons) $z_d$, which can be obtained by using a fitting
formula \cite{ref:Eisenstein}:
\begin{eqnarray}
&&z_d=\frac{1291(\Omega_mh^2)^{0.251}}{1+0.659(\Omega_mh^2)^{0.828}}[1+b_1(\Omega_bh^2)^{b_2}],
\end{eqnarray}
with
\begin{eqnarray}
&&b_1=0.313(\Omega_mh^2)^{-0.419}[1+0.607(\Omega_mh^2)^{0.674}], \\
&&b_2=0.238(\Omega_mh^2)^{0.223}.
\end{eqnarray}
In this paper, we use the data of $r_s(z_d)/D_V(z)$, which are
listed in Table \ref{baodata}, where $r_s(z)$ is the comoving sound
horizon size
\begin{eqnarray}
r_s(z)&=&c\int_0^t\frac{c_sdt}{a}=c\int_0^a\frac{c_sda}{a^2H}=c\int_z^\infty
dz\frac{c_s}{H(z)} \nonumber\\
&=&\frac{c}{\sqrt{3}}\int_0^{1/(1+z)}\frac{da}{a^2H(a)\sqrt{1+(3\Omega_b/(4\Omega_\gamma)a)}},
\end{eqnarray}
where $c_s$ is the sound speed of the photon$-$baryon fluid
\cite{ref:Hu1, ref:Hu2, ref:Caldwell}:
\begin{eqnarray}
&&c_s^{-2}=3+\frac{9}{4}\times\frac{\rho_b(z)}{\rho_\gamma(z)}=3+\frac{9}{4}\times(\frac{\Omega_b}{\Omega_\gamma})a,
\end{eqnarray}
and here $\Omega_\gamma=2.469\times10^{-5}h^{-2}$ for
$T_{CMB}=2.725K$.

\begin{table}[htbp]
\begin{center}
\begin{tabular}{c|l}
\hline\hline
 $z$ &\ $r_s(z_d)/D_V(z)$  \\ \hline
 $0.2$ &\ $0.1905\pm0.0061$  \\ \hline
 $0.35$  &\ $0.1097\pm0.0036$  \\
\hline
\end{tabular}
\end{center}
\caption{\label{baodata} The observational $r_s(z_d)/D_V(z)$
data~\cite{ref:Percival2}.}
\end{table}
Using the data of BAO in Table \ref{baodata} and the inverse
covariance matrix $V^{-1}$ in \cite{ref:Percival2}:

\begin{eqnarray}
&&V^{-1}= \left(
\begin{array}{cc}
 30124.1 & -17226.9 \\
 -17226.9 & 86976.6
\end{array}
\right).
\end{eqnarray}

The radial (line-of-sight) BAO scale measurement from galaxy power
spectra give constraint to cosmological parameters via the relation
\begin{equation}
\Delta z_{BAO}(z)=\frac{H(z)r_s(z_d)}{c}
\end{equation}
at two redshifts $z=0.24$ and $z=0.43$, the corresponding values are
$\Delta z_{BAO}(z=0.24)=0.0407\pm0.0011$ and $\Delta
z_{BAO}(z=0.43)=0.0442\pm0.0015$ respectively \cite{ref:BAOradial}.

Thus, the $\chi^2_{BAO}(p_s)$ is given as
\begin{equation}
\chi^2_{BAO}(p_s)=X^tV^{-1}X+\frac{[\Delta
z_{BAO}(z=0.24)-0.0407]^2}{0.0011^2}+\frac{[\Delta
z_{BAO}(z=0.43)-0.0442]^2}{0.0015^2},\label{eq:chi2BAO}
\end{equation}
where $X$ is a column vector formed from the values of theory minus
the corresponding observational data, with
\begin{eqnarray}
&&X= \left(
\begin{array}{c}
 \frac{r_s(z_d)}{D_V(0.2)}-0.1905 \\
 \frac{r_s(z_d)}{D_V(0.35)}-0.1097
\end{array}
\right),
\end{eqnarray}
and $X^t$ denotes its transpose.

\subsection{Cosmic Microwave Background constraints}

The CMB shift parameter $R$ is provided by \cite{ref:Bond1997}
\begin{equation}
R(z_{\ast})=\frac{\sqrt{\Omega_m
H^2_0}}{\sqrt{|\Omega_k|}}\mathrm{sinn}[\sqrt{|\Omega_k|}\int_0^{z{_\ast}}\frac{dz'}{H(z')}],
\end{equation}
here, the redshift $z_{\ast}$ (the decoupling epoch of photons) is
obtained by using the fitting function \cite{Hu:1995uz}
\begin{equation}
z_{\ast}=1048\left[1+0.00124(\Omega_bh^2)^{-0.738}\right]\left[1+g_1(\Omega_m
h^2)^{g_2}\right],\nonumber
\end{equation}
where the functions $g_1$ and $g_2$ read
\begin{eqnarray}
g_1&=&0.0783(\Omega_bh^2)^{-0.238}\left(1+ 39.5(\Omega_bh^2)^{0.763}\right)^{-1},\nonumber \\
g_2&=&0.560\left(1+ 21.1(\Omega_bh^2)^{1.81}\right)^{-1}.\nonumber
\end{eqnarray}
In addition, the acoustic scale is related to the distance ratio and
is expressed as
\begin{eqnarray}
&&l_A=\frac{\pi}{r_s(z_{\ast})}\frac{c}{\sqrt{|\Omega_k|}}\mathrm{sinn}[\sqrt{|\Omega_k|}\int_0^{z_\ast}\frac{dz'}{H(z')}].
\end{eqnarray}

\begin{table}[htbp]
\begin{center}
\begin{tabular}{c|ccc}
\hline\hline
  &\ $\mathrm{7-year}$ $\mathrm{ML}$ &\ $\mathrm{7-year}$ $\mathrm{mean}$ &\ $\mathrm{error}$, $\mathrm{\sigma}$ \\ \hline
 $l_A(z_\ast)$ &\ $302.09$ &\ $302.69$ &\ $0.76$ \\ \hline
 $R(z_\ast)$ &\ $1.725$ &\ $1.726$ &\ $0.018$ \\ \hline
 $z_{\ast}$  &\ $1091.3$ &\ $1091.36$ &\ $0.91$ \\
\hline
\end{tabular}
\end{center}
\caption{\label{cmbdata} The observational $l_A, R, z_{\ast}$
data~\cite{CMB:7yr}. The $\mathrm{ML}$ values are used in this work
as recommended.}
\end{table}

Using the data of $l_A, R, z_{\ast}$ in \cite{CMB:7yr}, which are
listed in Table \ref{cmbdata}, and their covariance matrix of
$[l_A(z_\ast), R(z_\ast), z_{\ast}]$ referring to \cite{CMB:7yr}:
\begin{eqnarray}
&&C^{-1}= \left(
\begin{array}{ccc}
2.305 & 29.698 & -1.333\\
 29.698 & 6825.270 & -113.180\\
 -1.333 & -113.180 & 3.414
\end{array}
\right),
\end{eqnarray}
we can calculate the likelihood $L$ as $\chi^2_{CMB}=-2\ln L$:
\begin{eqnarray}
&&\chi^2_{CMB}=\bigtriangleup d_i[C^{-1}(d_i,d_j)][\bigtriangleup
d_i]^t,
\end{eqnarray}
where $\bigtriangleup d_i=d_i-d_i^{data}$ is a row vector, and
$d_i=(l_A, R, z_{\ast})$.

\subsection{Gamma Ray Bursts}

Following \cite{ref:Schaefer}, we consider the well-known Amati's
$E_{p,i}-E_{iso}$ correlation \cite{ref:Amati'srelation,r16,r17,r18}
in GRBs, where $E_{p,i}=E_{p,obs}(1+z)$ is the cosmological
rest-frame spectral peak energy, and $E_{iso}$ is the isotropic
energy
\begin{equation}
E_{iso}=4\pi d^2_LS_{solo}/(1+z)
\end{equation}
in which $d_L$ and $S_{bolo}$ are the luminosity distance and the
bolometric fluence of the GRBs respectively. Following
\cite{ref:Schaefer}, we rewrite the Amati's relation as
\begin{equation}
\log\frac{E_{iso}}{{\rm erg}}=a+b\log\frac{E_{p,i}}{300{\rm
keV}}.\label{eq:calib}
\end{equation}

In \cite{ref:wang}, Wang defined a set of model-independent distance
measurements $\{\bar{r}_p(z_i)\}$:
\begin{equation}
\bar{r}_p(z_i)\equiv\frac{r_p(z)}{r_p(z_{0})},\quad r_p(z)\equiv
\frac{(1+z)^{1/2}}{z}\frac{H_0}{c}r(z),\label{eq:rp}
\end{equation}
where $r(z)=d_L(z)/(1+z)$ is the comoving distance at redshift $z$,
$z_{0}$ is the lowest GRBs redshift. Then, the cosmological model
can be constrained by GRBs via
\begin{eqnarray}
\chi^2_{GRBs}(p_s)&=&[\Delta\bar{r}_p(z_i)]\cdot(Cov^{-1}_{GRB})_{ij}\cdot[\Delta\bar{r}_p(z_i)],\label{eq:chi2GRB}\\
\Delta\bar{r}_p(z_i)&=&\bar{r}^{data}_p(z_i)-\bar{r}_p(z_i),
\end{eqnarray}
where $\bar{r}_p(z_i)$ is defined by Eq. (\ref{eq:rp}) and
$(Cov^{-1}_{GRB})_{ij},i,j=1...N$ is the covariance matrix. In this
way, the constraints from amount observational GRBs data are
projected into the relative few quantities $\bar{r}_p(z_i),i=1...N$.

Following the method proposed by Wang \cite{ref:wang}, Xu obtained
$N=5$ model-independent distances data points and their covariance
matrix by using $109$ GRBs via Amati's $E_{p,i}-E_{iso}$ correlation
\cite{ref:GRBsXu}. The resulted model-independent distances and
covariance matrix from $109$ GRBs are shown below in Tab.
\ref{tab:distance}
\begin{table}[htbp]
\begin{center}
\begin{tabular}{c|c|c|c|c}
\hline\hline
 & $z$ & $\bar{r}^{data}_p(z)$ & $\sigma(\bar{r}_p(z))^+$ &  $\sigma(\bar{r}_p(z))^-$\\ \hline
$0$ & $\quad0.0331\quad$ & $\quad1.0000\quad$ & $-$ & $-$  \\
$1$ & $1.0000$  & $0.9320$ & $0.1711$ & $0.1720$ \\
$2$ & $2.0700$  & $0.9180$ & $0.1720$ & $0.1718$ \\
$3$ & $3.0000$  & $0.7795$ & $0.1630$ & $0.1629$ \\
$4$ & $4.0480$  & $0.7652$ & $0.1936$ & $0.1939$ \\
$5$ & $8.1000$  & $1.1475$ & $0.4297$ & $0.4389$ \\
  \hline\hline
\end{tabular}
\end{center}
\caption{\label{Tab:datapoints} Distances measured form $109$ GRBs
via Amati's correlation with $1\sigma$ upper and lower uncertainties
\cite{ref:GRBsXu}. $z_{0}=0.0331$ as lowest redshift was
adopted.}\label{tab:distance}
\end{table}
and Eq. (\ref{eq:covM}). The $\{\bar{r}_p(z_i)\},i=1,...,5$
correlation matrix is given by
\begin{eqnarray}
&&(\overline{Cov}_{GRB})= \left(
\begin{array}{ccccc}
$1.0000$ & $0.7780$ & $0.8095$ & $0.6777$ & $0.4661$ \\
$0.7780$ & $1.0000$ & $0.7260$ & $0.6712$ & $0.3880$ \\
$0.8095$ & $0.7260$ & $1.0000$ & $0.6046$ & $0.5032$ \\
$0.6777$ & $0.6712$ & $0.6046$ & $1.0000$ & $0.1557$ \\
$0.4661$ & $0.3880$ & $0.5032$ & $0.1557$ & $1.0000$
\end{array}
\right),
\end{eqnarray}
and the covariance matrix is given by
\begin{equation}
(Cov_{GRB})_{ij}=\sigma(\bar{r}_p(z_i))\sigma(\bar{r}_p(z_j))(\overline{Cov}_{GRB})_{ij},\label{eq:covM}
\end{equation}
where
\begin{eqnarray}
\sigma(\bar{r}_p(z_i))=\sigma(\bar{r}_p(z_i))^+, \quad {\rm if}\quad
\bar{r}_p(z)\geq
\bar{r}_p(z)^{data}; \\
\sigma(\bar{r}_p(z_i))=\sigma(\bar{r}_p(z_i))^-, \quad {\rm if}\quad
\bar{r}_p(z)< \bar{r}_p(z)^{data},
\end{eqnarray}
the $\sigma(\bar{r}_p(z_i))^+$ and $\sigma(\bar{r}_p(z_i))^-$ are
the $1\sigma$ errors listed in Tab. \ref{Tab:datapoints}.

\subsection{The X-ray gas mass fraction constraints}
According to the X-ray cluster gas mass fraction observation, the
baryon mass fraction in clusters of galaxies (CBF) can be utilized
to constrain cosmological parameters. The X-ray gas mass fraction,
$f_{gas}$, is defined as the ratio of the X-ray gas mass to the
total mass of a cluster, which is approximately independent on the
redshift for the hot $(kT\gtrsim5keV)$, dynamically relaxed clusters
at the radii larger than the innermost core $r_{2500}$. The X-ray
gas mass fraction, $f_{gas}$, can be derived from the observed X-ray
surface brightness profile and the deprojected temperature profile
of X-ray gas under the assumptions of spherical symmetry and
hydrostatic equilibrium. Basing on these assumptions above, Allen et
al. \cite{ref:07060033} selected 42 hot $(kT\gtrsim5keV)$, X-ray
luminous, dynamically relaxed clusters for $f_{gas}$ measurements.
The stringent restriction to the selected sample can not only reduce
maximally the effect of the systematic scatter in $f_{gas}$ data,
but also ensure that the $f_{gas}$ data is independent on
temperature. In the framework of the $\Lambda \textmd{CDM}$
reference cosmology, the X-ray gas mass fraction is presented as
\cite{ref:07060033}
\begin{eqnarray}
&&f_{gas}^{\Lambda \textmd{CDM}}(z)=\frac{K A \gamma
b(z)}{1+s(z)}\left(\frac{\Omega_b}{\Omega_m}\right)
\left[\frac{d_A^{\Lambda \textmd{CDM}}(z)}{d_A(z)}\right]^{1.5},\ \
\ \ \label{eq:f_g}
\end{eqnarray}
where $A$ is the angular correction factor, which is caused by the
change in angle for the current test model $\theta_{2500}$ in
comparison with that of the reference cosmology
$\theta_{2500}^{\Lambda CDM}$:
\begin{eqnarray}
&&A=\left(\frac{\theta_{2500}^{\Lambda
\textmd{CDM}}}{\theta_{2500}}\right)^\eta \approx
\left(\frac{H(z)d_A(z)}{[H(z)d_A(z)]^{\Lambda
\textmd{CDM}}}\right)^\eta,
\end{eqnarray}
here, the index $\eta$ is the slope of the $f_{gas}(r/r_{2500})$
data within the radius $r_{2500}$, with the best-fit average value
$\eta=0.214\pm0.022$ \cite{ref:07060033}. And the angular diameter
distance is given by
\begin{eqnarray}
&&d_A(z)=\frac{c}{(1+z)\sqrt{|\Omega_k|}}\mathrm{sinn}[\sqrt{|\Omega_k|}\int_0^z\frac{dz'}{H(z')}],
\end{eqnarray}
where $\mathrm{sinnn}(\sqrt{|\Omega_k|}x)$ respectively denotes
$\sin(\sqrt{|\Omega_k|}x)$, $\sqrt{|\Omega_k|}x$,
$\sinh(\sqrt{|\Omega_k|}x)$ for $\Omega_k<0$, $\Omega_k=0$ and
$\Omega_k>0$.

In equation (\ref{eq:f_g}), the parameter $\gamma$ denotes
permissible departures from the assumption of hydrostatic
equilibrium, due to non-thermal pressure support; the bias factor
$b(z)= b_0(1+\alpha_b z)$ accounts for uncertainties in the cluster
depletion factor; $s(z)=s_0(1 +\alpha_s z)$ accounts for
uncertainties of the baryonic mass fraction in stars and a Gaussian
prior for $s_0$ is employed, with $s_0=(0.16\pm0.05)h_{70}^{0.5}$
\cite{ref:07060033}; the factor $K$ is used to describe the combined
effects of the residual uncertainties, such as the instrumental
calibration and certain X-ray modelling issues, and a Gaussian prior
for the 'calibration' factor is considered by $K=1.0\pm0.1$
\cite{ref:07060033};

Following the method in Ref. \cite{ref:CBFchi21,ref:07060033} and
adopting the updated 42 observational $f_{gas}$ data in Ref.
\cite{ref:07060033}, the best fit values of the model parameters for
the X-ray gas mass fraction analysis are determined by minimizing,
\begin{eqnarray}
&&\chi^2_{CBF}=\sum_i^N\frac{[f_{gas}^{\Lambda
\textmd{CDM}}(z_i)-f_{gas}(z_i)]^2}{\sigma_{f_{gas}}^2(z_i)},
\end{eqnarray}
where $\sigma_{f_{gas}}(z_i)$ is the statistical uncertainties
(Table 3 of \cite{ref:07060033}). As pointed out in
\cite{ref:07060033}, the acquiescent systematic uncertainties have
been considered according to the parameters i.e. $\eta, b(z), s(z)$
and $K$.

\subsection{Lookback Time}

Since the seminal work of Sandage \cite{ref:LT} who defines the lookback time as the difference
between the present age of the Universe ($t_0$) and its age ($t_z$)
when a particular light ray at redshift z was emitted, one has used the lookback time-redshift relation
to constrain cosmological models, for example lookback time as a constraint to DGP model \cite{ref:DGPZhu}.
The lookback time-redshift relation is given by
\begin{equation}
t_{L}(z)=\int^z_0 \frac{dz'}{(1+z')H(z')}.
\end{equation}
Following Capozziello et al. \cite{ref:Capozziello}, one can define the age $t(z_i)$ of an object (e.g., a galaxy, a quasar or a galaxy cluster)
at redshift $z_i$ as the difference between the age of the Universe at
$z_i$ and the age $z_F$ when the object was born,
\begin{eqnarray}
t(z_i)&=&\int^{\infty}_{z_i}\frac{dz'}{(1+z')H(z')}-\int^{\infty}_{z_F}\frac{dz'}{(1+z')H(z')}\nonumber\\
&=&t_{L}(z_F)-t_{L}(z_i).
\end{eqnarray}
Then the observed lookback time to an object at $z_i$ can be defined as
\begin{eqnarray}
t_{L}^{obs}(z_i)&=&t_{L}(z_F)-t(z_i)=[t^{obs}_0-t(z_i)]-[t^{obs}_0-t_{L}(z_F)]\nonumber\\
&=&t^{obs}_0-t(z_i)-df,
\end{eqnarray}
where $df$ is the delay factor which accounts for our ignorance
about the absolute age of universe when the object is formed at
$t(z_F)$. To constrain a cosmological model by using lookback time
of an object, one can use
\begin{equation}
\chi^2_{age}=\sum_i \frac{[t_L(z_i)-t^{obs}_L(z_i,df)]^2}{\sigma^2_T}+\frac{[t_0-t^{obs}_0]^2}{\sigma^2_{t^{obs}_0}}\label{eq:chi2age}
\end{equation}
where $\sigma^2_T=\sigma^2_i+\sigma^2_{t^{obs}_0}$, $\sigma_i$ is
the uncertainty of the individual lookback time to the $i^{th}$
galaxy of our sample and $\sigma_{t^{obs}_0}$ is the uncertainty on
the total expansion age of the universe. After marginalizing the
'nuisance' parameter $df$, one can use the following method to
constrain a cosmological model by using lookback time
\cite{ref:LTMarg}
\begin{eqnarray}
\chi^2_{LT}(p_s)&=&-2\ln \int^{\infty}_0 d(df) \exp(-\chi^2_{age}/2)\nonumber\\
&=&A-\frac{B^2}{C}+D-2\ln [\sqrt{\frac{\pi}{2C}}{\rm
erfc}(\frac{B}{\sqrt{2C}})],
\end{eqnarray}
where
\begin{equation}
A=\sum_i \frac{\Delta^2}{\sigma^2_T},\quad B=\sum_i \frac{\Delta}{\sigma^2_T},\quad C=\sum_i\frac{1}{\sigma^2_T},
\end{equation}
where $\Delta$ is
\begin{equation}
\Delta=t_L(z_i)-[t^{obs}_0-t(z_i)]
\end{equation}
and $D$ is the second term of Eq. (\ref{eq:chi2age}), ${\rm
erfc}(x)=1-{\rm erf(x)}$ is the complementary error function of the
variable $x$. The observational data points of the age of galaxies
are shown in Tab. \ref{Tab:LTdata} . The current observational
universe age is $t^{obs}_0=13.75\pm0.13$Gyr \cite{CMB:7yr}.

\begin{table}[htb]
\begin{center}
\begin{tabular}{c|c}
$z_i$ & $t_i(z_i)$ (Gyr)\\
\hline\hline
0.1171 & 10.2  \\
0.1174 &  10.0 \\
0.2220 &   9.0 \\
0.2311 &   9.0 \\
0.3559 &   7.6 \\
0.4520 &   6.8 \\
  0.5750 &  7.0\\
0.6440 &  6.0\\
0.6760 &  6.0\\
0.8330 &  6.0\\
0.8360 &  5.8\\
0.9220 &  5.5\\
1.179 &  4.6\\
1.222 & 3.5 \\
1.224 &   4.3 \\
1.225 &  3.5 \\
1.226 &  3.5 \\
1.340 &  3.4 \\
1.380 &  3.5 \\
1.383 &  3.5 \\
1.396 &  3.6 \\
  1.430 &  3.2 \\
1.450 &   3.2 \\
1.488 &  3.0 \\
1.490 &  3.6 \\
1.493 &  3.2 \\
1.510 &  2.8 \\
1.550 &  3.0 \\
1.576 &  2.5 \\
  1.642 &  3.0 \\
1.725 &    2.6 \\
1.845 & 2.5 \\
  0.60  &   9.20 \\
  0.70   &  9.80  \\
  0.80    & 3.41  \\
  0.10    & 3.08  \\
  0.25    & 4.84  \\
  1.27    & 12.13 \\
 \hline\hline
\end{tabular}
\end{center}
\caption{Galaxy ages \citet{svj}. The last $6$ data points are taken
from the Capozziello et al. \cite{ref:Capozziello} (Table 1) ages of
6 galaxy clusters in the redshift range $0.10 <  z < 1.27$. The
$1\sigma$ error are taken as $12\%$ of the age of galaxy with
exception of the last $6$ data points. The $1\sigma$ error of the
last $6$ data points are set as $1$.}\label{Tab:LTdata}
\end{table}

\subsection{Growth Function of Matter Linear Perturbations}

The linear growth factor of the matter density perturbation $D(a)$
is defined as
\begin{equation}
D(a)\equiv
\frac{\frac{\delta\rho}{\rho}(a)}{\frac{\delta\rho}{\rho}(a=1)},
\end{equation}
which is subject to the evolution equation
\cite{ref:CBFchi21,ref:Dequation}
\begin{equation}
D^{''}(k,a)+\left(\frac{3}{a}+\frac{E'(a)}{E(a)}\right)D^{'}(k,a)-\frac{3}{2}\frac{\Omega_{m0}}{a^5E(a)}f(k,a)D(k,a)=0
\end{equation}
with initial conditions $D(a)\simeq a$ for $a\simeq 0$ on sub-Hubble
scales, where $'$ denotes derivative with respect to scale factor
$a$ and $E(a)=H(a)/H_0$. The forms of $f(k,a)$ depend on the
dynamical equations of particular gravity theory. For general
relativity, $f(k,a)\equiv 1$. For the case of DGP model, $f(k,a)$
takes the form
\begin{equation}
f(k,a)=(1+\frac{1}{3\alpha}),
\end{equation}
where $\alpha$ is
\begin{equation}
\alpha=1-\frac{E(a)}{\sqrt{\Omega_{r_c}}}\left(1+\frac{a}{3}\frac{E'(a)}{E(a)}\right).
\end{equation}

The observational growth rate relates to $D(a)$ via
\begin{equation}
g(a)\equiv\frac{aD'(a)}{D(a)}=\frac{d\ln \delta}{d\ln a}.
\end{equation}

The current data sets about growth function are listed in Tab.
\ref{tab:fobs}.
 \begin{table}[htbp]
 \begin{center}
 \begin{tabular}{ccccc} \hline\hline
 $z_i$ &  ~~~~~$g^{obs}_i$ & Reference\\ \hline
 $0.15$ &~~~~~ $0.51\pm 0.11$ & \cite{hawk}\\
 $0.35$
 & ~~~~~$0.70\pm 0.18$ & \cite{teg}\\
 $0.55$ &~~~~~ $0.75\pm 0.18$ & \cite{ross}\\
 $0.77$ &~~~~~ $0.91\pm 0.36$ & \cite{guzz}\\
 $1.4$
 &~~~~~ $0.90\pm 0.24$ & \cite{angl}\\
 $3.0$ &~~~~~ $1.46\pm 0.29$ & \cite{dona}\\
 \hline\hline
 \end{tabular}
 \end{center}
 \caption{\label{tab:fobs} Observed perturbation growth as a function of redshift $z$, where $g(a)=\frac{d\ln \delta}{d\ln a}$ in terms of $a$.}
 \end{table}

Then, the constraint from growth function is in the form
\begin{equation}
\chi^2_{GF}(p_s)=\sum_i
\left[\frac{g(z_i)-g^{obs}_i}{\sigma_i}\right]^2.
\end{equation}


\begin{thebibliography}{*}

\bibitem{ref:Riess98} A. G. Riess {\it et al.}, Astron. J. {\bf 116} 1009 (1998) [astro-ph/9805201].

\bibitem{ref:Perlmuter99} S. Perlmutter {\it et al.}, Astrophys. J. {\bf 517} 565 (1999) [astro-ph/9812133].

\bibitem{ref:reviews} S.~Weinberg,
 {\it The cosmological constant problem,}
  Rev.\ Mod.\ Phys.\  {\bf 61}, 1 (1989);
V.~Sahni and A.~A.~Starobinsky,
  {\it The Case for a Positive Cosmological Lambda-term,}
  Int.\ J.\ Mod.\ Phys.\  D {\bf 9}, 373 (2000)
  [arXiv:astro-ph/9904398];
S.~M.~Carroll,
 {\it The cosmological constant,}
  Living Rev.\ Rel.\  {\bf 4}, 1 (2001)
  [arXiv:astro-ph/0004075];
P.~J.~E.~Peebles and B.~Ratra,
  {\it The cosmological constant and dark energy,}
  Rev.\ Mod.\ Phys.\  {\bf 75}, 559 (2003)
  [arXiv:astro-ph/0207347];
T.~Padmanabhan,
  {\it Cosmological constant: The weight of the vacuum,}
  Phys.\ Rept.\  {\bf 380}, 235 (2003)
  [arXiv:hep-th/0212290];
E.~J.~Copeland, M.~Sami and S.~Tsujikawa,
 {\it Dynamics of dark energy,}
  Int.\ J.\ Mod.\ Phys.\  D {\bf 15}, 1753 (2006)
  [arXiv:hep-th/0603057].

\bibitem{ref:DGP}  G. R. Dvali, G. Gabadadze, M. Porrati, Phys.
Lett. B. 485, 208(2000) [arXiv:hep-th/0005016].

\bibitem{ref:DGPreview} A. Lue, Phys. Rept. 423, 1(2006) [arXiv:astro-ph/0510068].

\bibitem{ref:DGPGc} M. Li, X. Li, X. Zhang, arXiv:0912.3988v1[astro-ph.CO].

\bibitem{ref:LDGPDc} H. Zhang, H. Yu, H. Noh, Z. Zhu, Phys. Lett. B 665,
319(2008) arXiv:0806.4082v2[astro-ph].

\bibitem{ref:DGPFang} L. Lombriser, W. Hu, W. Fang, U. Seljak, Phys. Rev. D
80,063536(2009).

\bibitem{ref:DGPwei} H. Wei, Phys. Lett. B 664,1(2008) arXiv:0802.4122v3
[astro-ph].

\bibitem{ref:gfweak} S. Nesseris and L. Perivolaropoulos, Phys. Rev. D 77, 023504 (2008)
[arXiv:0710.1092].

\bibitem{ref:DGPFE} C. Deffayet et al., Phys. Rev. D 66, 024019 (2002).

\bibitem{ref:MCMC} A. Lewis and S. Bridle, Phys. Rev. D {\bf 66} 103511 (2002);
URL: http://cosmologist.info/cosmomc/.


\bibitem{ref:07060033} S. W. Allen, D. A. Rapetti, R. W. Schmidt, H. Ebeling, R. G. Morris
and A. C. Fabian, Mon. Not. Roy. Astron. Soc. {\bf 383} 879 (2008).

\bibitem{ref:0409574} D. Rapetti, S. W. Allen and J. Weller, Mon. Not. Roy. Astron. Soc.
{\bf 360} 555 (2005).

\bibitem{ref:modifiedMCMC} URL: http://www.stanford.edu/\~ drapetti/fgas\_module/


\bibitem{ref:SN557} R. Amanullah et al. [Supernova Cosmology Project Collaboration], arXiv:1004.1711
[astro-ph.CO].

\bibitem{ref:SNchi2}
S. Nesseris and L. Perivolaropoulos, Phys. Rev. D {\bf  72} 123519
(2005); L. Perivolaropoulos, Phys. Rev. D {\bf 71} 063503 (2005); E.
Di Pietro and J. F. Claeskens, Mon. Not. Roy. Astron. Soc. {\bf 341}
1299 (2003); A.~C.~C.~Guimaraes, J.~V.~Cunha and J.~A.~S.~Lima, JCAP
{\bf 0910} 010 (2009).

\bibitem{ref:smallomega} E.~Garcia-Berro, E.~Gaztanaga, J.~Isern, O.~Benvenuto and
L.~Althaus, astro-ph/9907440; A. Riazuelo and J. Uzan, Phys. Rev. D
{\bf 66} 023525 (2002); V. Acquaviva and L. Verde, JCAP {\bf 0712}
001 (2007).


\bibitem{ref:Okumura2007}
T.~Okumura, T.~Matsubara, D.~J.~Eisenstein, I.~Kayo, C.~Hikage,
A.~S.~Szalay and D.~P.~Schneider, Astrophys. J.  {\bf 676} 889
(2008).

\bibitem{ref:Percival2} W. J. Percival {\it et al.}, arXiv:0907.1660 [astro-ph.CO].

\bibitem{ref:Eisenstein05} D. J. Eisenstein {\it et al.}, [SDSS Collabaration], Astrophys. J. {\bf 633} 560 (2005)
[astro-ph/0501171].

\bibitem{ref:Percival3} W. J. Percival {\it et al.}, Mon. Not. R. Astron. Soc. {\bf 381} 1053 (2007) arXiv:0705.3323 [astro-ph.CO].

\bibitem{ref:Eisenstein} D. J. Eisenstein and W. Hu, Astrophys. J. {\bf496} 605 (1998).


\bibitem{ref:Hu1} W. Hu and N. Sugiyama, Astrophys. J. {\bf 444} 489 (1995) [arXiv:astro-ph/9407093].

\bibitem{ref:Hu2} W. Hu, M. Fukugita, M. Zaldarriaga and M. Tegmark, Astrophys. J. {\bf 549} 669 (2001) [arXiv:astro-ph/0006436].

\bibitem{ref:Caldwell} R. R. Caldwell and M. Doran, Phys. Rev. D {\bf 69} 103517 (2004).

\bibitem{ref:BAOradial} E. Gaztanaga, R. Miquel, E. Sanchez, Phys. Rev. Lett. 103
091302(2009).


\bibitem{ref:Bond1997} J.~R.~Bond, G.~Efstathiou and M.~Tegmark, Mon. Not. Roy. Astron. Soc.  {\bf 291} L33 (1997).

\bibitem{Hu:1995uz} W.~Hu and N.~Sugiyama, Astrophys. J. {\bf 471} 542 (1996).

\bibitem{CMB:7yr} E. Komatsu et al. [WMAP Collaboration], arXiv:1001.4538 [astro-ph.CO].


\bibitem{ref:Schaefer} B. E. Schaefer, Astrophys. J. 660, 16 (2007) [astro-ph/0612285].

\bibitem{ref:Amati'srelation} L. Amati et al., Astron. Astrophys. 390, 81 (2002) [astro-ph/0205230].

\bibitem{r16}
L.~Amati {\it et al.},
 Mon.\ Not.\ Roy.\ Astron.\ Soc.\  {\bf 391}, 577 (2008)
 [arXiv:0805.0377].

\bibitem{r17}
L.~Amati, arXiv:1002.2232 [astro-ph.HE]; L. Amati, Mon. Not. Roy.
Astron. Soc. 372, 233 (2006) [astro-ph/0601553].

\bibitem{r18}
L.~Amati, F.~Frontera and C.~Guidorzi, arXiv:0907.0384
[astro-ph.HE].

\bibitem{ref:wang} Y. Wang, Phys.Rev.D 78,123532(2008).

\bibitem{ref:GRBsXu} L. Xu, arXiv:1005.5055v1 [astro-ph.CO].




\bibitem{ref:CBFchi21} S. Nesseris and L. Perivolaropoulos, JCAP {\bf 0701} 018
(2007) [astro-ph/0610092].




\bibitem{ref:LT} A. Sandage, Annu. Rev. Astron. Astrophys. 26, 561
(1988).

\bibitem{ref:DGPZhu} N. Pires, Z. Zhu, J. S. Alcaniz, Phys.Rev.D 73, 123530 (2006).

\bibitem{ref:Capozziello} S. Capozziello, V. F. Cardone, M. Funaro, and S. Andreon,
Phys. Rev. D 70, 123501 (2004).

\bibitem{ref:LTMarg} M. A. Dantas, J. S. Alcaniz, D. Jain, A.Dev, Astron. Astrophys. 467 (2007) 421.

\bibitem{svj} Simon, J., Verde, L., \& Jimenez, R.~2005, Phys. Rev. D, 71, 123001.

\bibitem{ref:Dequation} J. P. Uzan, Gen.Rel.Grav.39,307(2007)
[arXiv:astro-ph/0605313v1];  H. F. Stabenau, B. Jain, Phys.Rev. D74,
084007(2006) [arXiv:astro-ph/0604038v3].

 \bibitem{hawk}
E.~Hawkins {\it et al.},
 Mon.\ Not.\ Roy.\ Astron.\ Soc.\  {\bf 346}, 78 (2003)
 [astro-ph/0212375];\\
L.~Verde {\it et al.},
 Mon.\ Not.\ Roy.\ Astron.\ Soc.\  {\bf 335}, 432 (2002)
 [astro-ph/0112161];\\
E.~V.~Linder, arXiv:0709.1113 [astro-ph].
\bibitem{teg}
M.~Tegmark {\it et al.} [SDSS Collaboration],
 Phys.\ Rev.\  D {\bf 74}, 123507 (2006) [astro-ph/0608632].

\bibitem{ross}
N.~P.~Ross {\it et al.}, astro-ph/0612400.

\bibitem{guzz}
L.~Guzzo {\it et al.}, Nature\ {\bf 451}, 541 (2008)
 [arXiv:0802.1944].

\bibitem{angl}
J.~da Angela {\it et al.}, astro-ph/0612401.

\bibitem{dona}
P.~McDonald {\it et al.}  [SDSS Collaboration],
 Astrophys.\ J.\  {\bf 635}, 761 (2005) [astro-ph/0407377].
\end{thebibliography}
\end{document}